\documentclass[aps,prl,reprint,superscriptaddress]{revtex4-1}
\usepackage{graphicx}
\usepackage{dcolumn}
\usepackage{bm}
\usepackage{amsmath}

\usepackage{amsfonts}
\usepackage{textcomp} 

\usepackage{verbatim}
\usepackage {ulem}

\usepackage{color}
\usepackage{array}
\usepackage{makecell}

\newcolumntype{C}{>{\centering\arraybackslash$}m{2cm}<{$}}

\begin{document}

\title{CoFeAlB alloy with low damping and low magnetization for spin transfer torque switching}

\author{A. Conca}

\email{conca@physik.uni-kl.de}

\affiliation{Fachbereich Physik and Landesforschungszentrum OPTIMAS, Technische Universit\"at
Kaiserslautern, 67663 Kaiserslautern, Germany}

\author{T.~Nakano}

\affiliation{Department of Applied Physics, Tohoku University, Japan}

\author{T.~Meyer}

\affiliation{Fachbereich Physik and Landesforschungszentrum OPTIMAS, Technische Universit\"at
Kaiserslautern, 67663 Kaiserslautern, Germany}

\author{Y.~Ando}

\affiliation{Department of Applied Physics, Tohoku University, Japan}

\author{B.~Hillebrands}

\affiliation{Fachbereich Physik and Landesforschungszentrum OPTIMAS, Technische Universit\"at
Kaiserslautern, 67663 Kaiserslautern, Germany}

\date{\today}

\begin{abstract}
We investigate the effect of Al doping on the magnetic properties of the alloy CoFeB. Comparative measurements  of the  saturation magnetization, the Gilbert damping parameter $\alpha$ and the exchange constant as a function of the annealing temperature for CoFeB and CoFeAlB thin films are presented.
Our results reveal a strong reduction of the magnetization for CoFeAlB in comparison to CoFeB. If the prepared CoFeAlB films are amorphous, the damping parameter $\alpha$ is unaffected by the Al doping in comparison to the CoFeB alloy. In contrast, in the case of a crystalline CoFeAlB film, $\alpha$ is found to be reduced. Furthermore, the x-ray characterization and the evolution of the exchange constant with the annealing temperature indicate a similar crystallization process in both alloys. The data proves the suitability of CoFeAlB  for spin torque switching properties where a reduction of the switching current in comparison with CoFeB is expected.

\end{abstract}

\maketitle

The alloy CoFeB is widely used in magnetic tunneling junctions in combination with MgO barriers due to the large magnetoresistance effect originating in the spin filtering effect \cite{yuasa,yuasa2,choi,zhang}. For the application in magnetic random access memories, the switching of the magnetization of the free layer via  spin transfer torque (STT) with spin polarised currents is a key technology. However, the required currents for the switching process are still large and hinder the applicability of this technique. The critical switching current density for an in-plane magnetized system is given by \cite{diao} 

\begin{equation} \label{critical}
J_{\rm c0}= \frac{2e\alpha M_{\rm S} t_{\rm f} (H_{\rm K} + H_{\rm ext} + 2\pi M_{\rm S})}{\hbar \eta}
\end{equation}
where $e$ is the electron charge, $\alpha$ is the  Gilbert damping parameter, $M_{\rm S}$ is the saturation magnetization, $t_{\rm f}$ is the thickness of the free layer, $H_{\rm ext}$ is the external
field, $H_{\rm K}$ is the effective anisotropy field  and $\eta$ is the spin transfer efficiency. From the expression  it is clear that, concerning material parameters, $J_{\rm c0}$ is ruled by the product $\alpha M_{\rm S}^2$. For out-of-plane oriented layers, the term $2\pi M_{\rm S}$ vanishes and then $J_{\rm C0}$ is proportional to $\alpha M_{\rm S}$ \cite{wang}. Even in the case of using pure spin currents created by the Spin Hall effect, the required currents are proportional to factors of the form $\alpha^n M_{\rm S}$ with $n=1,1/2$ \cite{taniguchi}. A proper strategy to reduce the critical switching currents is then defined by reducing the saturation magnetization. This can be achieved by the development of new materials or the modification of known materials with promising properties.  Since the compatibility with a MgO tunneling barrier and the spin filtering effect must be guaranteed together with industrial applicability, the second option is clearly an advantage by reducing  $M_{\rm S}$ in the CoFeB alloy. In this case, a critical point  is that this reduction must not be associated with an increase of the damping parameter $\alpha$.

\begin{figure}[b]
    \includegraphics[width=0.9\columnwidth]{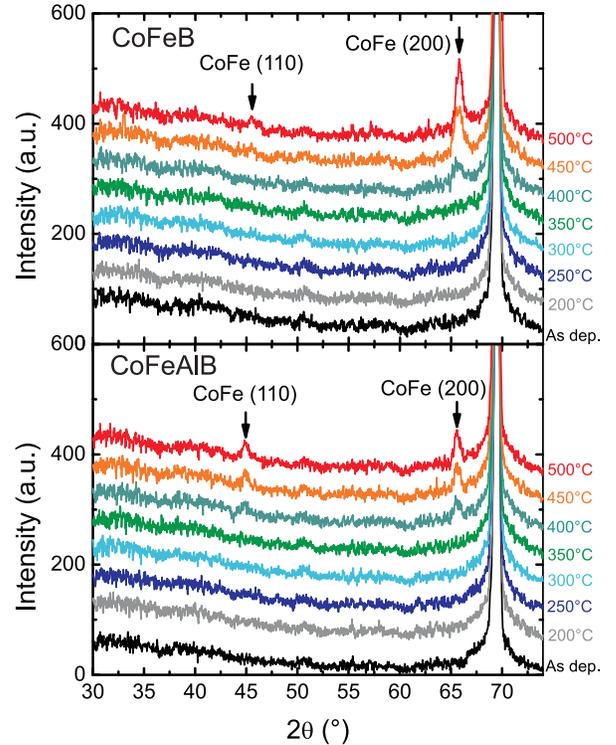}
	  \caption{\label{xrays}(Color online) $\theta/2\theta$-scans for 40~nm thick films of Co$_{40}$Fe$_{40}$B$_{20}$ (top) and Co$_{36}$Fe$_{36}$Al$_{18}$B$_{10}$ (bottom)  showing the evolution of crystallization with the annealing temperature.}
\end{figure}

In the last years, several reports on doped CoFeB alloys have proven the potential of this approach. The introduction of Cr results in a strong reduction of $M_{\rm S}$ \cite{oguz,kubota,cui}, however, it is sometimes also causing an increase of the damping parameter \cite{oguz}. The reduction of $M_{\rm S}$ by doping CoFeB with Ni is smaller compared to a doping with Cr but it additionally leads to a  reduction of $\alpha$ \cite{oguz}. In constrast, the reduction of magnetization with V is comparable to Cr \cite{kubota} but to our knowledge no values for $\alpha$ have been published.  In the case of doping of CoFeB by Cr or by V, a reduction of the switching current has been shown \cite{oguz,kubota}. 

In this Letter, we report on results on Al doped CoFeB alloy thin films characterized by ferromagnetic resonance spectroscopy. The dependence of $M_{\rm S}$, the Gilbert damping parameter $\alpha$ and the exchange constant on the annealing temperature is discussed together with the crystalline structure of the films and the suitability for STT switching devices.

The samples are grown on Si/SiO$_2$ substrates using  DC (for metals) and RF (for MgO) sputtering techniques. The layer stack of the samples is Si/SiO$_2$/Ta(5)/MgO(2)/FM(40)/MgO(2)/Ta(5) where FM = Co$_{40}$Fe$_{40}$B$_{20}$ (CoFeB) or Co$_{36}$Fe$_{36}$Al$_{18}$B$_{10}$ (CoFeAlB). Here, the values in brackets denote the layer thicknesses in nm. In particular, the FM/MgO interface is chosen since it is widely used for STT devices based on MTJs. This interface is also required to promote the correct crystallization of the CoFeB layer upon annealing since the MgO layer acts as a template for a  CoFe bcc (100)-oriented structure \cite{yuasa,yuasa2,choi} with consequent B migration.

\begin{figure}[t]
    \includegraphics[width=0.9\columnwidth]{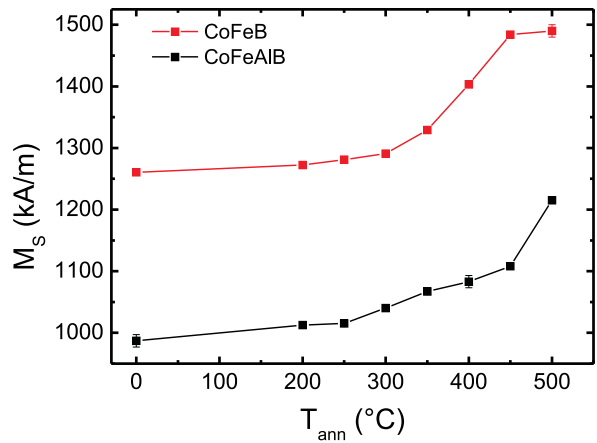}
	  \caption{\label{meff}(Color online) Evolution of the saturation magnetization for CoFeB and CoFeAlB with the annealing temperature T$_{\rm ann}$.}
\end{figure}

The dynamic properties and material parameters were studied by measuring the ferromagnetic resonance using a strip-line vector network analyzer (VNA-FMR). For this, the samples were placed face down and the S$_{12}$ transmission parameter was recorded. A more detailed description of the FMR measurement and
analysis procedure is shown in previous work \cite{fept,cofeb-ann}. Brillouin light spectroscopy (BLS) was additionally used for the measurement of the exchange constant. The crystalline  bulk properties of  the films were studied by X-ray diffractometry (XRD) using the Cu-K$_\alpha$ line.

\begin{figure}[t]
    \includegraphics[width=0.9\columnwidth]{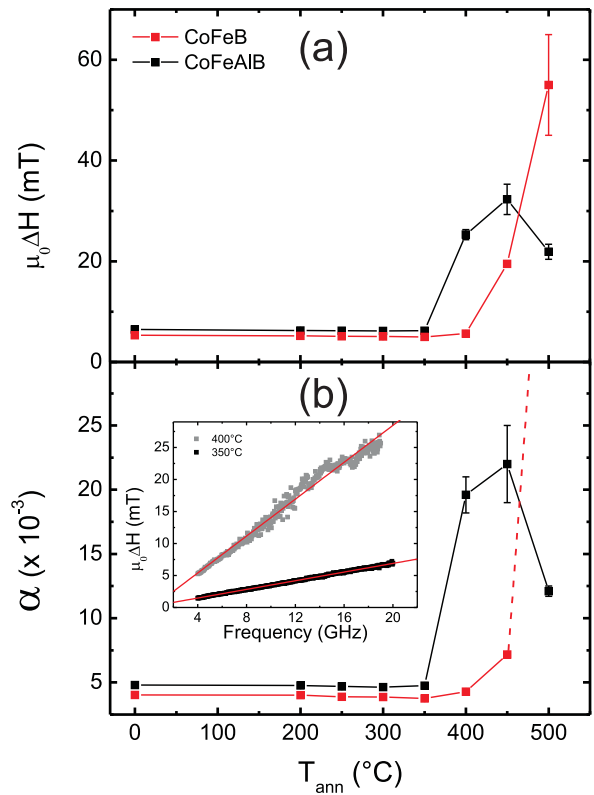}
	  \caption{\label{damping}(Color online) Linewidth at a fixed frequency of 18 GHz (a) and Gilbert damping parameter $\alpha$  dependence on the annealing temperature T$_{\rm ann}$ (b). The $\alpha$ value for $T_{\rm ann} =500^{\circ}$  is only a rough estimation since the large linewidth value does not allow for a proper estimation. The inset shows the linear dependence of the linewidth on the frequency exemplarily for CoFeAlB annealed at 350$^{\circ}$C and 400$^{\circ}$C. The red lines are a linear fit. }
\end{figure}

Figure~\ref{xrays} shows the $\theta/2\theta$-scans for CoFeB (top) and CoFeAlB (bottom) samples annealed at different temperatures T$_{\rm ann}$. The appearance of the CoFe diffractions peaks, as shown by the arrows in Fig.~\ref{xrays} indicate the start of crystallization  at high annealing temperatures of more than 400$^{\circ}$C. In the case of lower annealing temperatures or the as-deposited samples, the FM layer is in an amorphous state. The first appearance of the (200) diffraction peak occurs at the same point for both alloys showing a very similar thermal evolution. This simplifies a substitution of CoFeB by the Al alloy in tunneling junctions since the same annealing recipes can be applied. This is critical since the used values must be also optimized for the quality of the tunneling barrier itself or  the perpendicular anisotropy induced by the FM/MgO interface.  The (110) CoFe peak is also present for both material compositions owing to a partial texturing of the film. However, the larger intensity of the (200) peak is not compatible with a random crystallite orientation but with a dominant (100) oriented film \cite{concas,you}. This is needed since the spin filtering effect responsible for the large magnetoresistance effect in MgO-based junctions requires a (100) orientation.

The dependence of the FMR frequency on the external magnetic field is described by Kittel's formula \cite{kittel}.
The value of $M_{\rm eff}$ extracted from the Kittel fit is related with the saturation magnetization of the sample and the interfacial properties by $M_{\rm eff}= M_{\rm S}-2K^{\perp}_{\rm S}/\mu_0 M_{\rm S}d $ where $K^{\perp}_{\rm S}$ is the interface  perpendicular anisotropy constant. For the thickness used in this work (40~nm) and physically reasonable $K^{\perp}_{\rm S}$ values, the influence of the interface is negligible and therefore $M_{\rm eff}\approx M_{\rm S}$. For details about the estimation of $M_{\rm eff}$ the reader is referred to \cite{cofeb-ann}.

Figure~\ref{meff} shows the obtained values for $M_{\rm S}$ for all samples. A strong reduction for CoFeAlB in comparison with standard CoFeB is observed and the relative difference is maintained for all T$_{\rm ann}$. The evolution with annealing is very similar for both alloys. Significantly, the increase in $M_{\rm S}$ starts for values of T$_{\rm ann}$ lower than expected  from the appearance of the characteristic    CoFe diffraction peaks in the XRD data (see Fig.~\ref{xrays}). This shows that the measurement of $M_{\rm S}$ is  the more sensitive method to probe the change of the crystalline structure.

For CoFeB a saturation value around $M_{\rm S}\approx$1500~kA/m is reached at T$_{\rm ann}=450^{\circ}$C. This is compatible with values reported for CoFe (1350-1700~kA/m) \cite{liu,bilzer} and CoFeB (1350-1500~kA/m) \cite{bilzer,liu2}. On the contrary, for CoFeAlB the introduction of Al reduces the magnetization of the samples and the annealing does not recover to CoFe-like values.

\begin{figure}[t]
    \includegraphics[width=0.9\columnwidth]{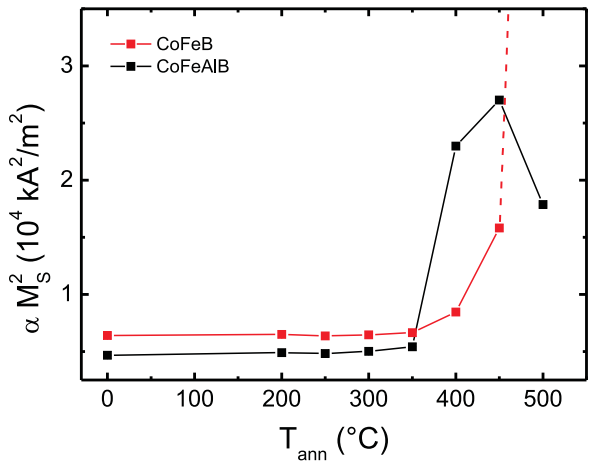}
	  \caption{\label{alphams2}(Color online) Dependence of the product $\alpha M_{\rm S}^2$ on the annealing temperature T$_{\rm ann}$ for CoFeB and CoFeAlB. This quantity is ruling the switching current in in-plane magnetized STT devices as shown in Eq.~\ref{critical}.}
\end{figure}

Figure~\ref{damping}(a) shows the dependence of the magnetic field linewidth on  T$_{\rm ann}$ measured at a fixed frequency of 18~GHz. From the linear dependence of this linewidth on the FMR frequency,  the Gilbert  damping parameter is extracted (as exemplarily shown for the CoFeAlB alloy in the inset in Fig.~\ref{damping}(b)) and the results are shown in Fig.~\ref{damping}(b). For T$_{\rm ann}$ values up to 350$^{\circ}$C, where the amorphous phase is still dominating, almost no difference between both alloys is observed. With increasing temperature the damping increases for both alloys but the evolution is different. For CoFeAlB the increase starts almost abruptly at T$_{\rm ann}=400^{\circ}$C, reaches a maximum around $\alpha=0.02$ and then decreases again to $\alpha=0.012$ for T$_{\rm ann}=500^{\circ}$C. In contrast,   the increase for CoFeB is more smoothly with T$_{\rm ann}$ and increases stadily with higher T$_{\rm ann}$. In fact, due to the large linewidths reached for T$_{\rm ann}=500^{\circ}$C, the value of $\alpha$ cannot be properly estimated and only a lower limit of 0.03-0.04 can be given. This situation is represented by the dashed line in Fig.~\ref{damping}(b). It is important to note here that when the crystallization process is fulfilled (i.e. for T$_{\rm ann}=500^{\circ}$C) $\alpha$ is much lower for the Al doped alloy. This is relevant for the application in tunneling junctions where a full crystallization is required for the presence of the spin filtering effect originating large magnetoresistance values in combination with MgO barriers \cite{zhang}.

For further comparison of both alloys, the quantity $\alpha M_{\rm S}^2$ has been calculated and plotted in Fig.~\ref{alphams2}. As shown in Eq.~\ref{critical}, this value is ruling  the critical switching current in in-plane magnetized systems. We observe for the alloys showing a mostly amorphous phase (T$_{\rm ann}<400^{\circ}$C) a slight improvement for CoFeAlB in comparison with CoFeB due  to the lower $M_{\rm S}$. However, for fully crystalline films (T$_{\rm ann}=500^{\circ}$C), the CoFeAlB shows a much smaller value for $\alpha M_{\rm S}^2$. Since a full crystalline phase is needed for any application of this alloy in MTJ-based devices, this denotes a major advantage of this compound compared to standard. 

\begin{figure}[t]
    \includegraphics[width=0.9\columnwidth]{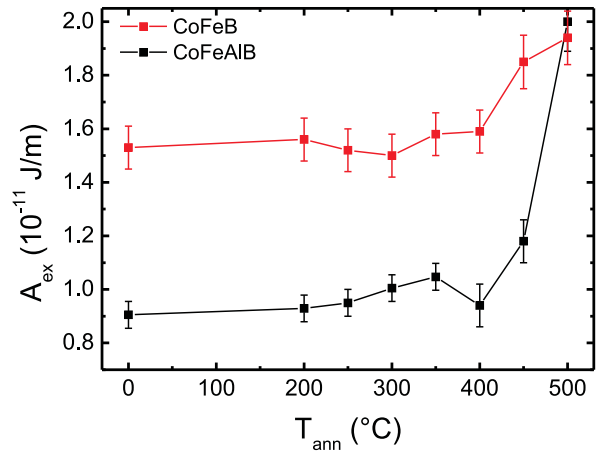}
	  \caption{\label{aex}(Color online) Dependence of the exchange constant $A_{\rm ex}$ on the annealing temperature T$_{\rm ann}$ for CoFeB and CoFeAlB. The top panels show typical BLS spectra  for materials (see text).}
\end{figure}

The exchange constant $A_{\rm ex}$ is a critical parameter that is strongly influenced by the introduction of Al. Its estimation in required for modeling the spin torque switching behavior of the alloys. The access to the constant is given by the dependence of the frequency of the perpendicular standing spin-wave (PSSW) modes on the external static magnetic field \cite{demokritov}. As shown in previous works \cite{conca,cofeb-ann}, it is possible to observe the PSSW modes in metallic films with a standard VNA-FMR setup. However, the signal is strongly reduced  compared to the FMR peak. For the samples presented in this paper,   the PSSW peak could not be observed for T$_{\rm ann}>400^{\circ}$C since the increased damping leads to a broadening and lowering of the peak which prevents the estimation of $A_{\rm ex}$. For this reason, BLS spectroscopy is  used for the measurement of the frequency position of the PSSW modes. This technique has a larger sensitivity for the PSSW modes than VNA-FMR. 

Figure \ref{aex}(c) shows the evolution of $A_{\rm ex}$ upon annealing for both alloys. For the films dominated by the amorphous phase the value is much lower for CoFeAlB which is also compatible with the lower magnetization. However, as the crystallization evolves, the exchange constant increases stronger than for CoFeB and the same value is obtained for the fully crystallized films. This fact points to a similar role of Al and B during the crystallization process: when the CoFe crystallites form, the light atoms are expelled forming a Al-B-rich matrix embedding the magnetic crystallites. This explains also the similar evolution observed in the XRD data shown in Fig.~\ref{xrays}. The lower maximal magnetization obtained for the CoFeAlB can be explained by the reduced CoFe content but also a certain number of residual Al and B atoms  in the crystallites, which may differ for both alloys.

The $A_{\rm ex}$ values for as-deposited CoFeB films are very similar to previous reports \cite{conca,cofeb-ann,cho}. Concerning the values for the crystallized samples, since the properties are strongly dependent on the B content and of the ratio between Co and Fe as well as on the exact annealing conditions, a comparison with literature has to be made carefully. Nevertheless, the maximal value  and the evolution with T$_{\rm ann}$ for CoFeB is similar to the one reported by some of the authors \cite{cofeb-ann}. Also results for alloys with the same B content are compatible with our data \cite{helmer,sato}. CoFeB films with reduced B content show  larger values \cite{bilzer}, the same is true for CoFe alloys with values between 3.84-2.61$\times10^{11}$~J/m depending on the exact stoichiometry \cite{liu,bilzer}. This may again be a hint that a rest of Al or B is present in the CoFe crystallites.

In summary, the presented experimental results show that CoFeAlB is a good candidate as alternative to CoFeB for spin torque switching devices due to the reduction of the factor $\alpha M_{\rm S}^2$ which dominates the critical switching current. This reduction was found to originate from a strong  reduction of the saturation magnetization  and a decreased damping parameter $\alpha$ for fully crystalline CoFeAlB films. Furthermore, the results reveal a larger thermal stability of the damping properties  in CoFeAlB compared to CoFeB. The absolute values of $M_{\rm S}$ and the exchange constant $A_{\rm ex}$ for crystalline films point to a formation of CoFe crystallines with a non-vanishing content of the lights atoms embedded in a B or Al matrix.

Financial support by  M-era.Net through the HEUMEM project, the DFG in the framework of the research unit TRR
173  Spin+X and by the JSPS Core-to-Core Program is gratefully acknowledged.


\begin{thebibliography}{sotief}

\bibitem{yuasa} S.~Yuasa and D.~D.~Djayaprawira, J.~Phys.~D: Appl.~Phys. {\bf 40}, R337-R354 (2007).

\bibitem{yuasa2} S.~Yuasa,Y.~Suzuki, T.~Katayama, and K.~Ando, Appl.~Phys.~Lett. \textbf{87}, 242503 (2005).

\bibitem{choi} Y.~S.~Choi, K.~Tsunekawa, Y.~Nagamine, and D.~Djayaprawira, J.~Appl.~Phys. \textbf{101}, 013907 (2007).

\bibitem{zhang}X.-G.~Zhang and W.~H.~Butler, J.~Phys.: Condens.~Matter \textbf{15} R1603, (2003).

\bibitem{diao} Z.~Diao, Z.~Li, S.~Wang, Y.~Ding, A.~Panchula, E.~Chen, L.-C.~Wang, and Y.~Huai, J.~Phys.~D: Appl.~Phys. \textbf{19}, 165209 (2007). 

\bibitem{wang} K.~L.~Wang, J.~G.~Alzate, and P.K.~Amiri, J.~Phys.~D: Appl.~Phys. \textbf{46}, 074003 (2013). 

\bibitem{taniguchi} T.~Taniguchi, S.~Mitani, and M.~Hayashi, Phys.~Rev.~B \textbf{92}, 024428 (2015).
	
\bibitem{oguz} K.~Oguz, M.~Ozdemir, O.~Dur, and J.~M.~D.~Coey, J.~Appl.~Phys. \textbf{111}, 113904 (2012).

\bibitem{kubota} H.~Kubota, A.~Fukushima, K.~Yakushiji, S.~Yakata, S.~Yuasa, K.~Ando, M.~Ogane, Y.~Ando, and T.~Miyazaki, J.~Appl.~Phys. \textbf{105}, 07D117 (2009).

\bibitem{cui} Y.~Cui, M.~Ding, S.~J.~Poon, T.~P.~Adl, S.~Keshavarz, T.~Mewes, S.~A.~Wolf, and J.~Lu, J.~Appl.~Phys. \textbf{114}, 153902 (2013).

\bibitem{fept} A.~Conca, S.~Keller, L.~Mihalceanu, T.~Kehagias, G.~P.~Dimitrakopulos, B.~Hillebrands,  and E.~Th.~Papaioannou,  Phys.~Rev.~B {\bf 93}, 134405 (2016).

\bibitem{cofeb-ann} A.~Conca, E.~Th.~Papaioannou, S.~Klingler, J.~Greser, T.~Sebastian, B.~Leven, J.~L\"osch, and B.~Hillebrands,  Appl.~Phys.~Lett. {\bf 104}, 182407 (2014).

\bibitem{concas} G.~Concas, F.~Congiu, G.~Ennas, G.~Piccaluga, and G.~Spano. J.~of Non-Crystalline Solids \textbf{330}, 234 (2003).
\bibitem{you} C.~Y.~You, T.~Ohkubo, Y.~K.~Takahashi, and K.~Hono, J.~Appl.~Phys. \textbf{104}, 033517 (2008).

\bibitem{kittel} C.~Kittel,  Phys. Rev. {\bf 73}, 155 (1948).

\bibitem{liu} X.~Liu, R.~Sooryakumar, C.~J.~Gutierrez, and G.~A.~Prinz, J. Appl. Phys. \textbf{75}, 7021 (1994).
	
\bibitem{bilzer} C.~Bilzer, T.~Devolder, J.-V.~Kim, G.~Counil, C.~Chappert, S.~Cardoso, and P.~P.~Freitas, J. Appl. Phys. \textbf{100}, 053903 (2006).

\bibitem{liu2} X.~Liu, W.~Zhang, M.~J.~Carter, and G.~Xiao, J. Appl. Phys. \textbf{110}, 033910 (2011).

\bibitem{demokritov}S.~O.~Demokritov,  B.~Hillebrands, {\it Spin Dynamics in Confined Magnetic
Structures I}, Springer, Berlin, (2002).


\bibitem{conca}A.~Conca, J.~Greser, T.~Sebastian, S.~Klingler, B.~Obry, B.~Leven, and B.~Hillebrands,  J.~Appl.~Phys. \textbf{113}, 213909 (2013).


\bibitem{cho} J.~Cho, J.~Jung, K.-E.Kim, S.-I.~Kim, S.-Y.~Park, and M.-H.~Jung, C.-Y.~You, J. of Magn and Magn. Mat. \textbf{339}, 36 (2013).


\bibitem{helmer} A.~Helmer, S.~Cornelissen, T.~Devolder, J.-V.~Kim, W.~van Roy, L.~Lagae, and C.~Chappert, Phys.~Rev.~B \textbf{81}, 094416 (2010).

\bibitem{sato} H.~Sato, M.~Yamanouchi, K.~Miura, S.~Ikeda, R.~Koizumi, F.~Matsukura, and H.~Ohno, IEEE Magn. Lett., \textbf{3}, 3000204 (2012).











\end{thebibliography}
\end{document}